\UseRawInputEncoding
\PassOptionsToPackage{english}{babel}
\documentclass[%
 amsmath,amssymb,
 aps,
prl,
reprint,
]{revtex4-2}

\usepackage{graphicx}
\usepackage{dcolumn}
\usepackage{bm}

\usepackage[english]{babel}
\usepackage{braket}
\usepackage{array}
\usepackage[colorinlistoftodos,prependcaption,textsize=tiny]{todonotes} 
\newcommand{\note}[1]{\textcolor{black}{#1}} 
\newcommand{\dotkp}{\boldsymbol{k}\cdot\boldsymbol{p}}
\usepackage{url}
\usepackage{makecell}

\begin{document}
\selectlanguage{English}

\preprint{APS/123-QED}

\title{\note{Terahertz Lasing at Roomtemperature: Numerical Study of a Vertical-Emitting} Quantum Cascade Laser Based on a Quantum Dot Superlattice}

\author{Alexander~Mittelst{\"a}dt}
 \email{mittelstaedt@tu-berlin.de}
\author{Ludwig~A.~Th.~Greif}%
\author{Stefan~T.~Jagsch}%
\author{Andrei~Schliwa}%
\affiliation{%
 Institut f{\"u}r Festk{\"o}rperphysik, Technische Universit{\"a}t Berlin, \\
 Hardenbergstr. 36, 10623 Berlin, Germany
}%

\date{\today}

\begin{abstract}
We investigate room temperature lasing of terahertz quantum cascade lasers using quantum dot chains as active material \note{suitable for wireless communication and imaging technologies.}
Bandstructure calculations for such extended systems of coupled quantum dots are made possible by a novel \lq linear combination of quantum dot orbitals\rq --method, based on single quantum dot wavefunctions.
\note{Our results demonstrate strong vertical-emission of coupled quantum dots}, reduced phonon coupling and in-plane scattering, enabling room-temperature lasing with significantly reduced threshold current densities.
%
\end{abstract}

\maketitle


%
%
In conventional quantum cascade lasers (QCLs), electrons run down a staircase potential generated by a superlattice of coupled quantum wells, where amplification of radiation occurs via electronic intra-band transitions, a concept first proposed by Kazarinov and Suris in 1971\ \cite{kazarinov_possibility_1971}.
Population inversion between the sub-bands is achieved by meticulous engineering of electron lifetimes and transition probabilities by means of layer thicknesses and external bias, thus, tuning intra-band transitions with meV accuracy.
\note{
Since the first realization of a QCL operating in the mid-infrared by Faist et al.\ in 1994\ \cite{faist_quantum_1994}, constant development of device design and material growth paved the way for the first QCL 
operating within the terahertz (THz) spectrum 
in 2002 \cite{kohler_terahertz_2002} and resulted in an operating temperature of up to 250\,K\ \cite{khalatpour_high-power_2020}.
Compact THz-QCL devices are of major interest for wireless communication applications, providing the high frequencies needed to meet the increasing demand for bandwidth\ \cite{nagatsuma_advances_2016, khalatpour_high-power_2020, sarieddeen_next_2020, miles_terahertz_2007}.
Also, as biological tissue and other materials are transparent for THz radiation, THz-QCLs are promising alternative light sources for non-invasive inspection and imaging \cite{williams_terahertz_2007, tonouchi_cutting-edge_2007, redo-sanchez_terahertz_2016-1}.
%
%
}
However, operation at room temperature is still impeded by an increasing competition from non-radiative scattering losses and free-carrier absorption\ \cite{ferreira_evaluation_1989, williams_terahertz_2007, kumar_recent_2011}, as well as a challenging carrier injection for transitions at energies below the material's longitudinal optical phonon frequency ($\sim$4$\,$--$\,$20$\,$meV).
Population inversion is further hindered by the continuous in-plane spectrum of quantum wells (QW), non-radiative scattering of carriers out of the upper laser level, and thermal backfilling of the lower laser level. As non-radiative decay rates are orders of magnitude larger than the radiative decay rates \cite{faist_wallplug_2007, faist_quantum_1994, faugeras_high-power_2005, evans_investigation_2006}, threshold current densities found in QCLs are generally high ($\sim$\,kAcm$^{-2}$), regardless of the operating wavelength.\\
To address these obstacles, QCLs with an active region composed of quantum dot (QD) chains have been proposed (QD-QCL) \cite{luryi_prospects_1996, dmitriev_quantum_2005}, where the localized states in the QDs lead to reduced electron-phonon scattering (phonon bottleneck) and free-carrier absorption, increasing carrier lifetimes by orders of magnitude and thereby improving temperature stability \cite{li_phonon_1999, zibik_intraband_2004, zibik_long_2009}. 
QD-QCLs also benefit from the intrinsically narrow gain spectrum of QD-chains, and significantly reduced threshold current densities are predicted \cite{dmitriev_quantum_2005, wingreen_quantum-dot_1997, chia-fu_hsu_intersubband_2000, vukmirovic_electron_2008}.
\begin{figure}[!t]
 	\centering
 	 \includegraphics[]{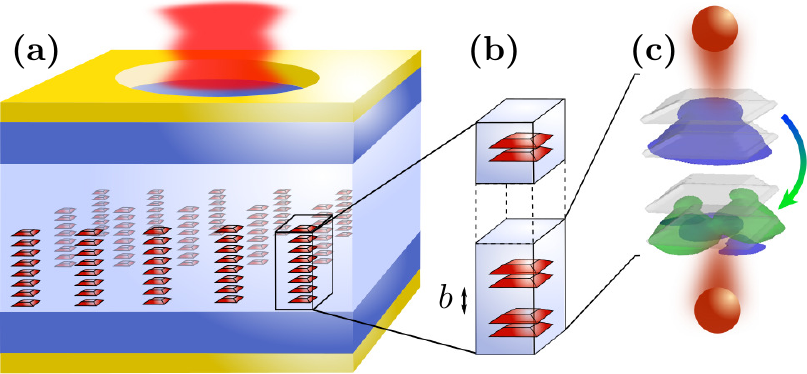} 
 	\caption{Schematics of a QD-QCL. (a) Chains of electronically coupled QDs form an array in the active region of the QCL. 
 	%
 	%
 	(b) QD-chain, with barriers of width $b$ separating the QD unit-cells.
 	%
    (c) Vertical emission of $s$-to-$p$-like transitions in stacked QDs.
 	}
 	\label{fig:fig_1}
\end{figure}%
\note{%
Additionally, QW-QCLs operate at comparatively high voltages since tunneling barriers between neighboring QWs have to be small to establish a superlattice.
In contrast, by using QD chains, the inter-dot distances can be larger, and therefore the bias can be smaller, resulting in reduced parasitic tunneling into higher or bulk states. 
Another advantage of QDs is their pronounced spatial emission anisotropy enabling active region designs based on vertically emitting $s$-to-$p$-like transitions such as vertical-cavity surface-emitting lasers:
then QD-based THz-QCLs also benefit from simple coupling into optical fibers, operation in arrays, and very low-cost production\ \cite{curwen_broadband_2019, czyszanowski_quantum-cascade_2018, shang_room-temperature_2017, mei_quantum_2017-1, bewley_room-temperature_2016, kneissl_vertical_2020}.\\
%
%
}%
In this letter, we present a bandstructure calculation for a stack of 20 QDs, featuring an intra-band staircase potential suitable for THz operation, \note{as well as a transport calculation.}
Our QD-QCL design comprises an array of InGaAs QD-chains with a two-QD unit-cell superlattice embedded in a GaAs matrix, cf.\ Fig.\ \ref{fig:fig_1}.
Strong electronic inter-dot coupling results in delocalized electronic states along the QD-chain, which are engineered to facilitate population inversion at a certain external bias.
Based on realistic device parameters, we find a strongly reduced threshold current density compared to QW-based heterostructures.
The InGaAs/GaAs material system provides both a mature growth platform and superior material quality\ \cite{detz_evaluation_2018, belkin_new_2015, williams_terahertz_2007}.\\
The challenge of finding suitable parameters for the QD gain material boils down to the computational cost of simulating i) stacks of more than ten coupled QDs as well as ii) calculating dozens of excited electronic states of these QD-chains.
QD-chain length, tunneling barrier width, and composition and geometry of the individual QDs are varied to find the desired bandstructure.
\note{
Since, so far, a rigorous simulation of the electronic structure of a realistic stack of QDs is still missing and to drastically reduce the associated computational cost, we developed a novel \lq linear combination of quantum dot orbitals\rq --method (LCQO), based on realistic single QD single-particle wavefunctions.
%
%
The method is not limited to calculating the electronic structure of extended QD systems but also any other three-dimensional configuration of coupled QDs, including all bound states, resulting in a realistic representation of the entire electronic system.
}
Our LCQO-model can be implemented on top of any atomistic or continuum model for the electronic states of the QDs and is in no way limited to the 8-band $\dotkp$\,-wavefunctions exemplarily used herein.
The LCQO-method is introduced, along with a detailed study of the electronic properties of stacks of coupled QDs, in (the associated manuscript) Ref.\ [PRB]. 
%
%
\section{Method of calculation}
The underlying idea of the LCQO-method is to split an extensive system of coupled QDs, i.e., a large eigenvalue problem, into a set of subsystems of single QDs for which single-particle states can be efficiently calculated.
\note{
Provided that $\ket{\varphi}$ denote the eigenstates of the QD subsystems, which build a composite basis set, this basis is then used to expand the eigenstates of the QD-chain system (QDC),
$\ket{\psi_{i}^{\textrm{QDC}}} = \sum_k a_{ik} \ket{\varphi_{k}}$, 
where $k$ indicates the single QD states.
The generalized eigenvalue problem
$
\sum \big[\Braket{\varphi_{l}|\boldsymbol{H}|\varphi_{k}} - \varepsilon_{i}  \Braket{\varphi_{l} | \varphi_{k}} \big]a_{ik} = 0 ,
\label{eq:gev}
$
yields eigenvalues $\varepsilon_{i}$ and coefficients $a_{ik}$, resulting in a set of LCQO single-particle functions $\ket{\psi_{i}^{\textrm{QDC}}}$.
}
Above, $\boldsymbol{H}$ denotes the Hamiltonian of the composite system, where we consider the strain distribution and resulting piezoelectric fields in the whole heterostructure.
To make the LCQO efficient, first, a library of single QDs varying in geometry and material composition is created, whose electronic single-particle states are subsequently calculated to serve as a basis in the LCQO calculations.
The crucial benefit of this method is the reduction to a variational problem for finding the stationary points for $\braket{\boldsymbol{H}}$ of the large-scale system as a function of a finite set of coefficients, whereby the calculation time is reduced by at least three orders of magnitude, compared to a full 8-band $\dotkp$ calculation.
A detailed derivation of the LCQO-method, a performance test, and a direct comparison to full $\bm{k}\cdot\bm{p}$ simulations are provided in (the associated manuscript) Ref.\ [PRB].\\
%
\note{
Transport within the QD-QCL system is modeled via a rate equation model, modified from \cite{koechner_solid-state_2006, michalzik_vcsels_2013},
}
\begin{align}
\note{\dot{N_{i}}} &= \note{ 
                                \left( \eta \frac{j(t)}{e \, a} \right)
                                + \sum_{i \neq j}^{} R_{j,i} - \sum_{i \neq j}^{} R_{i,j}
                                \mp \sum_{i \neq j}^{} R_{i,j}^{\textrm{pt}} 
                                }
                                \label{eq:N3}\\
\note{
\dot{S}} &= \note{
\sum_{i \neq j}^{} R_{i,j}^{\textrm{ind}} + \beta \sum_{i \neq j}^{} R_{i,j}^{\textrm{sp}} - \kappa S ,
}
\label{eq:S}
\end{align}
\note{where in Eq.\ \ref{eq:N3}, $R_{i,j}^{\textrm{pt}}$
is the sum of induced and spontaneous emission rates, relevant if $i$ denotes an upper ($-$) or lower laser level ($+$).
}
The rate equations for the carrier densities $N_{i}$ are connected to the occupation probabilities $n_{i}$ via $N_{i}=N_{p}^{*}n_{i}$, where $\note{N_{p}^{*} = 2 \rho / a }$, with the lateral density of QD-chains in the active region $\rho$ \note{and the length of a QD-QCL cascade $a$}.
\note{
The indices $i$ and $j$ comprise not simply the QD-chains delocalized QDC eigenstates, but also QD wetting layers (wl) and the bulk material (B), modeled using 2-D and 3-D carrier densities, respectively.}
The injection current density is $j(t)$, $e$ is the elementary charge, \note{and $\eta$ is the injection efficiency which shifts the threshold current linearly.
We assume a perfect funneling of carriers, i.e., $j(t)$ is considered only for the first unit-cell of the QD-chain.
}
\note{
At the QD-chains edge, tunneling of carriers into a quasi-metallic lead 
is modeled with a tunneling matrix element of $0.5$\,meV, c.f.\ Refs.\ \cite{goldozian_transport_2016-1, sprekeler_coulomb_2004}.
}\\
The non-radiative in- and out-scattering rates are defined as $R_{i,j}^{}=N_{i}(1-n_{j}) / \tau_{i,j}$,
where the relaxation times $\tau_{i,j}$
\note{
are calculated via a polaron-decay model adapted from Refs.\ \cite{zibik_long_2009, grange_polaron_2007, verzelen_polaron_2000}.
We assume that the transition energies between $s$-like and $p$-like states derived via pump-probe experiments on InGaAs QDs in Refs.\ \cite{zibik_long_2009, zibik_intraband_2004}, correspond to the energies $E_{i,j}$ calculated within our LCQO-model.
}%
\note{%
Additionally, we scale the phonon scattering by a linear dependence on the envelope-function overlap, 
c.f.\ Ref.\ \cite{vurgaftman_effect_1994, li_phonon_1999}.
}
%
Carrier escape and relaxation times are related via a quasi-Fermi distribution at 300\,K, 
\note{
and we scale the escape times for bound QD states into the bulk via a thermally assisted tunneling model,
cf.\ Refs.\ \cite{martin_electric_1981, schramm_thermionic_2009}.
Transport within the bulk is modeled via carrier drift in a strong electric field, c.f.\ Refs.\ \cite{blakemore_semiconducting_1982, rode_chapter_1975}.
}
\\
%
\note{
In Eq.\ \ref{eq:S}, $S$ is the photon density and the induced emission term is $R_{i,j}^{\textrm{ind}}$.
As the inhomogeneous size dispersion of the QDs within the QD-chains result in a broadening of the optical transitions, the probability of a transition having the energy $E_{i,j}$ is modeled by the Gaussian function 
analogous to the broadening of the QD states.
Additionally, each subgroup of optical transitions is represented by a Lorentzian function 
whereby 
lifetime and thermal broadening are considered.
%
%
$R_{i,j}^{\textrm{ind}}$ and the total cavity losses $\kappa$ in Eq.\ \ref{eq:S} are calculated with respect to the edge and vertical-emitting cavity designs, c.f.\ Refs.\ \cite{verdeyen_laser_1995, blood_quantum_2015, bosco_thermoelectrically_2019, williams_terahertz_2003, moser_energy-efficient_2015, basso_basset_spectral_2019, borri_ultralong_2001, kuntz_modulated_2006, bimberg_quantum_2003}. 
}
\\
The spontaneous emission term is $R_{i,j}^{\textrm{sp}}$, 
where the lifetime is determined from the QDC bandstructure via Fermi's golden rule.
\note{
Furthermore, we calculate the spatially resolved emission intensity of QDC intraband transitions as introduced in Ref.\ \cite{greif_tuning_2018}.
}
The $\beta$-factor defines the fraction of spontaneous emission coupled into the laser mode.
See the Supplemental Material at [URL] for details of the transport model and Refs.\ \cite{madelung_numerical_1982, burenkov1973temperature} for additional parameters.
\section{Discussion and Results}
\begin{figure}[t]
	\centering
	\includegraphics[]{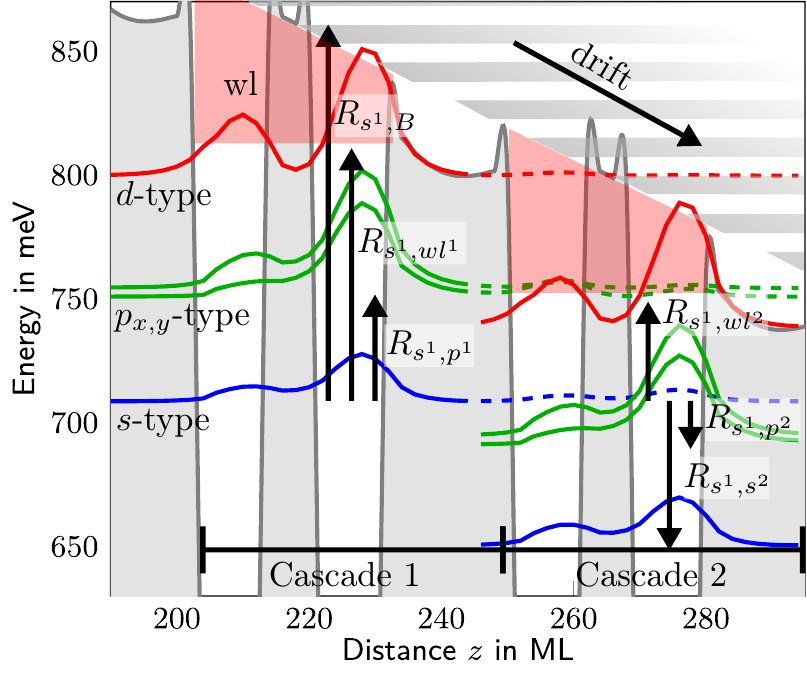}%
	\caption{
	\note{
	A snippet of the conduction band staircase potential for a stack of 20 InGaAs QDs with an alternating barrier sequence of $8$ and $20$\,MLs (grey shaded).
 	The electronic QDC single-particle states are shown, whereby the states and energy levels considered are indicated by blue ($s$-type), green ($p_{x,y}$-type), and red ($d$-type) lines.
 	The bandstructure shows two cascades at an applied bias of $46$\,kVcm$^{-1}$.
 	Arrows depict the transition rates $R_{i,j}$ exemplarily for an electron located in the upper two-QD unit-cell's $s$-type orbital.
 	Red shaded areas indicate the wetting layers and the hatched area represents the bulk.
 	%
 	The diagonal laser transition is realized between the upper two-QD unit-cell's $s$- and the neighboring two-QD unit-cell's $p_{x,y}$-type orbitals, i.e., $s^{1} \rightarrow p^{2}$.
 	The transition results in a photon energy of $\hbar \omega \approx 16\,$meV ($\approx3.8$\,THz).
 	%
 	%
	See the Supplemental Material at [URL] for the full bandstructure staircase potential.
	}
	}
	\label{fig:fig_2}
\end{figure}%
The realization of a THz QD-QCL imposes specific requirements on the design of the bandstructure. 
In particular, these are (i) cascaded intra-band transitions of identical energy with electron probability densities shared by neighboring QDs and (ii) in- and out-scattering rates that allow population inversion between the laser levels. 
Facilitated by the LCQO-method and based on the results and discussions in (the associated manuscript) Ref.\ [PRB], we initially studied more than a hundred possible realizations of QD-chains at various external biases, with the prerequisite of using experimentally realistic parameters.
The QDs are electronically coupled across tunneling barriers of width $b$, forming QD-chains, see Fig.\ \ref{fig:fig_1}(b).
The individual QD-chains are electronically uncoupled through a lateral separation much larger than the exciton Bohr radius of the material system, i.e., distance $\gg 15\,\mathrm{nm}$ for GaAs \cite{ekardt_determination_1979}.
The QD-chains consist of InGaAs QDs modeled as truncated pyramids, following TEM imaging of Stranski-Krastanov QDs \cite{blank_quantification_2009, litvinov_influence_2008}.
For the bandstructure presented in Fig.\ \ref{fig:fig_2}, we assumed In$_{0.7}$Ga$_{0.3}$As QDs with a side-wall inclination of $40^{\circ}$, a basis length of $14.7\,$nm and a height of $2.8\,$nm, in agreement with reports in \cite{yamauchi_electronic_2006,lemaitre_composition_2004, bruls_stacked_2003, sugaya_multi-stacked_2011}.
The vertical aspect ratio (height divided by base diameter) of $AR_{v}=0.135$ is chosen slightly lower to account for material interdiffusion found in experiments.
The finite difference grid's resolution for the LCQO calculations is set to two monolayers (MLs) of GaAs ($5.653\,$\AA).
Instead of a uniform QD separation, we use a two-QD unit-cell with QDs coupled through a $b=8$\,MLs barrier, with subsequent unit-cells separated by $b=20$\,MLs, which leads to a more readily achievable population inversion for the characteristic transition.
\\
\note{
Fig.\ \ref{fig:fig_2} shows a segment of the bandstructure of a stack of 20 QDs, which results in seven cascades along the QD-chain, c.f.\ the Supplemental Material at [URL].
Here, we trace the intra-dot and inter-dot transitions considered within Eqs.\ (\ref{eq:N3} --\ \ref{eq:S}) starting from a carrier located in the $s$-type orbital of the first cascade.
In general, such transitions, as well as the reverse transitions, are considered for each QDC eigenstate within the QD cascade and all cascades within the QD-chain, resulting in at least 35 coupled rate equations.
See the Supplemental Material for the approximations to the QD-QCLs bandstructure implemented to our transport model, reducing the computational cost.
}\\
As discussed in (the associated manuscript) Ref.\ [PRB], edge effects are essentially converged for chains of more than 10 QDs, after which the number of QCL cascades increases with subsequent stacking.
%
The laser transition is realized between the ground state ($s$-type symmetry) of a two-QD unit-cell and the neighboring unit-cell's excited state ($p$-type), see Fig.\ \ref{fig:fig_2}.
Symmetries of the orbitals are as specified in \cite{schliwa_impact_2007} and (the associated manuscript) Ref.\ [PRB].
\note{
%
We assign QDC states to a two-QD unit-cell, respectively to a cascade, whenever they have a significant local density, c.f.\ Fig.\ \ref{fig:fig_2}.
%
Here, we developed an active region inspired by designs used already successfully in QW-based QCLs working at record temperatures, c.f.\ Refs.\ \cite{kumar_1_8-thz_2011, kumar_recent_2011, bosco_thermoelectrically_2019, franckie_two-well_2018}.
\begin{figure}[!t]
	\centering
	\includegraphics[]{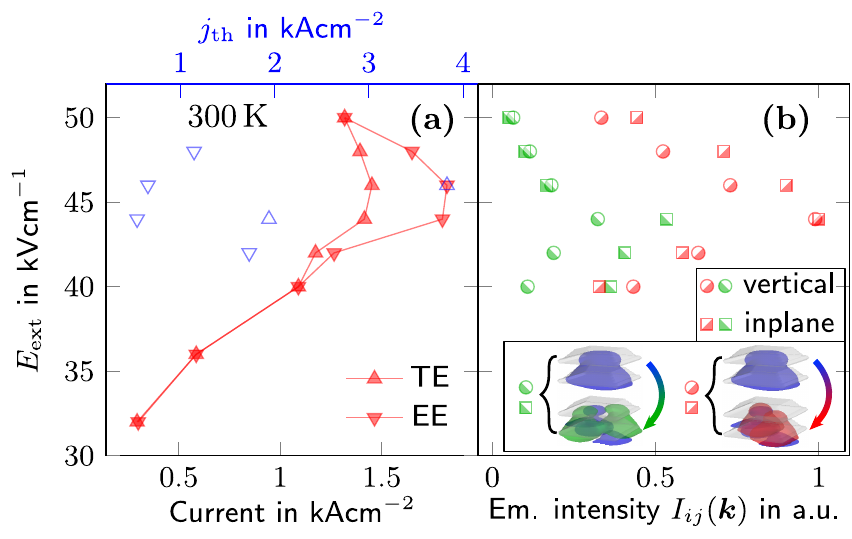}%
	\caption{
	\note{
	The current-voltage characteristics and the corresponding lateral and vertical emission intensity.
	(a) Current from the QD-chains into the lead as a function of the applied external field $E_{\textrm{ext}}$
	for the TE and EE designs (red filled triangles), at $j=2.2$\,kAcm$^{-2}$.
	Empty triangles depict the corresponding calculated threshold current densities (blue).
	(b) The emission intensity averaged over all lasing transitions within the QD-chain versus $E_{\textrm{ext}}$.
	Considering the emission intensity for the three perpendicular propagation directions, we show the vertical and the average inplane emission intensity, $ I_{i,j}(\textrm{vertical}) = I(\bm k = [001]) $ and $ I_{i,j}(\textrm{inplane}) = [ I(\bm k =[110]) + I(\bm k =[1\bar{1}0]) ] /2 $, respectively.
	%
	%
	%
	%
	$s$-to-$p_y$ and $s$-to-$p_x$ transitions are indicated by green and red semi-filled circles and squares, respectively.
	}
	}
	\label{fig:fig_3}
\end{figure}%
Our design facilitates population inversion as long as scattering out of the upper laser level is less efficient than depletion of the lower laser level into the QDs ground state.
%
This setting is always maintained for a specific bias interval since the strongly coupled QDs enable rapid relaxation of carriers within the two-QD unit-cell due to the considerable overlap of the eigenstates (intra-dot transition). 
However, the overlap of the states involved in transitions between the unit-cells is always smaller (inter-dot transition); see Supplemental Material for details.
The energy spacing of $E_{p,s}=44$\,meV results in an efficient intra-dot relaxation process by also reducing thermal backfilling of the lower laser level, c.f.\ Ref.\ \cite{bosco_thermoelectrically_2019}, and a reduced non-radiative scattering out of the upper laser level.
In contrast to the laser transition, the intra-dot $p$-to-$s$ relaxation energy is barely influenced by external bias but mainly determined by the material composition and the width of the coupled QDs barrier $b$, cf.\ (the associated manuscript) Ref.\ [PRB].
Thermally-induced losses due to scattering and tunneling out of the upper laser level are reduced compared to the lower laser level, as the $s$-type and $p$-type levels are ground and excited QD states, respectively.
}
As the ground state is the next period's upper laser level, the structure facilitates population inversion in analogy to reduced THz-QCL designs based on QWs, omitting an additional injector region \cite{ulbrich_intersubband_2002, kumar_186_2009, scalari_broadband_2010-1, bosco_thermoelectrically_2019} to maximize the number of QCL cascades, and therefore the gain, per QD-chain.
\note{
As our active region design facilitates a diagonal laser transition, its energy can be uniformly adjusted from $\sim$ 8\,--\,17\,meV, e.g., for the edge-emitter cavity design, using an external field of 42\,--\,48\,kVcm$^{-1}$.
}\\
%
%
We analyze transport through the device and the temperature-dependent threshold current density \note{
using two sets of realistic parameters for the top- and edge-emitter cavity designs (TE and EE); see Tab.\ \ref{tab:annahmen} for the parameters used.
%
%
%
Herein, the designs were optimized for a lasing frequency of $\hbar \omega \approx 16\,$meV ($\approx3.8$\,THz).
Fig.\ \ref{fig:fig_3}(a) shows a plot of the current coupled into the lead versus the applied external field.
%
%
}
\begin{table}
\caption{\label{tab:annahmen}Parameters used for the rate equation model.}
\begin{ruledtabular}
\begin{tabular}{cccc}
parameter & physical meaning & value & Ref.\\ 
  \hline
  $\rho$ & lateral density of QD-chains & 25e10\,cm$^{-2}$ \\
  $a$ & period length & 13.6\,nm \\
  $C$ & Num. of cascades per chain& 7 \\
  $m_{e}^{*}$ & eff. el. mass for In$_{0.7}$Ga$_{0.3}$As & $0.0356m_e$ &\cite{kasap_iii-v_2017}\\
  $m_\textrm{B}^{*}$ & eff. el. mass for GaAs & $0.063m_e$ &\cite{kasap_iii-v_2017}\\
  $\beta$ & beta-factor & 10$^{-4}$ \\
  $n_{\lambda}$ & mode index & 3.6 \\
  $\alpha_w$ & TE/EE tot. waveguide losses & 3/10\,cm$^{-1}$ & \cite{kuntz_modulated_2006, bimberg_quantum_2003} \\
  $l^{\textrm{EE}}$ & EE cavity length & 1\,mm \\
  $w^{\textrm{EE}}$ & EE cavity width & 180\,nm \\
  $r^{\textrm{EE}}$ & EE mirror reflectivity & 0.32 \\
  $r^{\textrm{TE}}$ & TE mirror reflectivity & 0.997 & \\
  $\gamma$ & FWHM of the QD ensemble & 8\,meV \\
  $\Lambda(T)$ & \makecell{FWHM for a QD sub-group\\ at $T=300$\,K} & 9\,meV &\cite{basso_basset_spectral_2019, borri_ultralong_2001} \\
\end{tabular}
\end{ruledtabular}
\end{table}%
%
%
\note{
Furthermore, our design makes it possible to continuously vary matrix elements and transition rates via external bias and the barrier width between unit-cells. At the same time, the size and composition of the QDs remain constant.
Stacking identical QDs would be considerably easier than the epitaxial growth of multiple types of QDs.
Using the two-QD unit-cells results in an improved localization of electron densities, thereby reducing the overlap with charge carriers within the neighboring unit-cells, increasing the upper laser level's lifetime.
\\
%
In Fig.\ \ref{fig:fig_3}(b), we show the normalized emission directionality for the $s^{1}$-to-$p_y^{2}$ (green) and $s^{1}$-to-$p_x^{2}$ (red) transitions versus the applied external bias.
%
Both transitions show a strong dependence of $I_{i,j}(\bm k)$ on the bias voltage and spatial anisotropy of the emission intensity, enabling QD-based QCLs in TE and EE devices.\\
}%
\begin{figure}[tb]
 	\centering
 	 \includegraphics[]{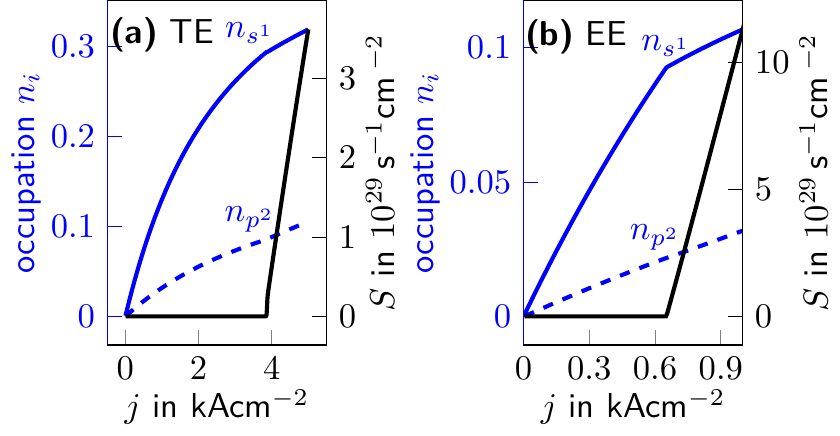} 
 	\caption{
 	\note{
 	The occupation probabilities (solid and dashed blue lines) for the upper ($s^{1}$) and lower laser levels ($p^{2}$) and the photon flux densities $S$ (solid black line) as a function of pump current density $j$ for the TE (a) and EE (b) laser designs, respectively.
 	The external field is set to $E_{\textrm{ext}}=46\,$kVcm$^{-1}$, and the temperature is set to $T=300$\,K.
 	}
 	}
 	\label{fig:fig_4}
\end{figure}%
\note{
Fig.\ \ref{fig:fig_4} depicts the occupation probabilities for the lasing transition, $s^{1} \rightarrow p^{2}$ in Fig.\ \ref{fig:fig_2}, and the corresponding photon flux density versus the pump current at 300\,K.
%
}
%
%
%
%
%
\note{
As discussed in (the associated manuscript) Ref.\ [PRB], minor fluctuations in geometry of the QDs within the stacks and a composition gradient show a much lower impact on intra-band compared to inter-band transitions and are not detrimental for the cascading and gain, c.f.\ also Refs.\ \cite{schliwa_impact_2007, fafard_coupled_2000}.
Considering a size dispersion of InGaAs QDs of $\sim$10\%, see, e.g., Ref.\ \cite{basso_basset_spectral_2019} and its impact on the QDs intra-band transition energies, we choose a full-width-half-maximum (FWHM) of 8\,meV for the QD ensemble's inhomogeneous broadening.
For the EE design at $E_{\textrm{ext}}=\note{46}\,$kVcm$^{-1}$, the threshold current density at 300\,K is $j_{th} \approx 660\,$Acm$^{-2}$, see Fig.\ \ref{fig:fig_3}(a), which is one order of magnitude lower than the threshold current density reported in Ref.\ \cite{khalatpour_high-power_2020, bosco_thermoelectrically_2019} for a quantum-well THz QCL operating at $>210$\,K.
Due to the reduced gain, the TE design results in $j_{th}\approx 3.8\,$kAcm$^{-2}$, significantly larger than the EE design.
}\\
%
%
\note{
In conclusion, we investigated room temperature lasing of vertical- or edge-emitting THz QCLs using QD-chains as an active material, a crucial next step  in developing devices suitable for wireless communication and imaging technologies.
}
\note{
Based on a novel LCQO approach, we are able to optimize the gain material efficiently and show a cascaded lasing transition within the THz spectral region at $\sim 3.8$\,THz.
Our two-QD unit-cell design allows a diagonal laser transition that is adjustable by external bias while maintaining a fast depletion of the lower laser level.
Furthermore, we show that optical matrix elements, transition rates, and the emission directionality can be uniformly tuned via the bias voltage and barrier widths between neighboring QD unit-cells.
The crucial aspect of achieving room-temperature operation is the strong suppression of non-radiative carrier losses and thermal backfilling by the quasi-zero-dimensional nature of the QDs.
}
\note{
Finally, based on extensive electronic structure calculations, our transport model predicts significantly reduced threshold current densities than QW-based QCLs, reducing heat dispersion.
}
Our results serve as a valuable guideline for future experiments towards room temperature THz QCLs.
\begin{acknowledgments}

The Deutsche Forschungsgemeinschaft partly funded this work in the framework of the SFB 787. We thank Markus R. Wagner and Dirk Ziemann for discussions and reading the manuscript.
\end{acknowledgments}
\bibliography{lib_korr}

\end{document}